\documentclass[aps,prl,twocolumn,groupedaddress]{revtex4-1}

\usepackage{amssymb}
\usepackage{graphicx}
\usepackage{dcolumn}
\usepackage{bm}
\usepackage{hyperref}
\usepackage{url}
\begin{document}

\title{Determination of neutrino masses from future observations of CMB B-mode polarization and the growth of structure}

\author{Koichi Hirano}
\email[Electronic address: ]{k\_hirano@tsuru.ac.jp}
\affiliation{Department of Elementary Education, Tsuru University, Tsuru 402-8555, Japan}

\date{\today}

\begin{abstract}
Constraints on neutrino masses are estimated based on future observations of the cosmic microwave background (CMB) including the B-mode polarization produced by CMB lensing using the Planck satellite, and baryon acoustic oscillations distance scale and the galaxy power spectrum from all-sky galaxy redshift survey in the BigBOSS experiment. We estimate the error in the bound on the total neutrino mass to be $\Delta\sum m_{\nu} = 0.012$ eV with a 68\% confidence level. If the fiducial value of the total neutrino mass is $\sum m_{\nu}= 0.06$ eV, this result implies that the neutrino mass hierarchy must be normal.
\end{abstract}


\maketitle

The standard model of particle physics assumes that neutrinos are massless. However, neutrino oscillation experiments indicate that neutrinos have nonzero masses. Experimental mass differences between the neutrinos are $|\Delta m_{21}^2| = 7.59^{+0.19}_{-0.21}\times 10^{-5}{\rm eV}^2$ \cite{aharmim2008} and $|\Delta m_{32}^2| = 2.43^{+0.13}_{-0.13}\times 10^{-3}{\rm eV}^2$ \cite{adamson2008}. However, the absolute masses and hierarchical structure have not yet been determined. These are essential information to build physics beyond the standard models.

Terrestrial experiments such as tritium beta decay \cite{thummler2011} and neutrinoless double-beta decay \cite{Gomez-Cadenas2011} give upper bounds on the absolute neutrino masses. Cosmological observations could further constrain neutrino properties by providing a more stringent bound on the total neutrino mass $\sum m_{\nu}$ and the effective number of neutrino species $N_\nu$.

Cosmic microwave background (CMB) anisotropies are generated mainly until eras before the last scattering surface of the decoupling epoch ($z \sim 1089$). Therefore, if neutrinos are as massive as $\sum m_{\nu} \hspace{0.3em}\raisebox{0.4ex}{$>$}\hspace{-0.75em}\raisebox{-.7ex}{$\sim$}\hspace{0.3em} 1.5$ eV, they become nonrelativistic before the recombination epoch. If so, a finite-mass neutrino significantly  affects the CMB spectrum. For masses below $\sum m_{\nu} \hspace{0.3em}\raisebox{0.4ex}{$<$}\hspace{-0.75em}\raisebox{-.7ex}{$\sim$}\hspace{0.3em} 1.5$ eV, the neutrinos alter the CMB spectrum chiefly through their effect on the angular diameter distance out to the last-scattering surface. The effect is degenerate with other cosmological parameters such as the matter energy density parameter $\Omega_{m0}$ and the Hubble constant $h$ \cite{ichikawa2005}. In that case, other cosmological probes complementary to CMB are needed to break the parameter degeneracy in order to study the small mass scales of neutrinos.

The B-mode polarization due to CMB lensing provides detailed information. It has better sensitivity to neutrino masses smaller than $0.1$ eV. This resolution is indispensable to distinguish between a normal and an inverted hierarchy.

The main effects of massive neutrinos on the growth of matter density perturbations arise from two physical mechanisms \cite{lesgourgues2006a}. First, a massive neutrino becomes nonrelativistic at the transition temperature, and contributes to the energy density of cold dark matter. That changes the matter-radiation equality time and the expansion rate of the universe. Second, the matter density perturbations are suppressed at small scales by neutrino free-streaming. Neutrinos travel at the speed of light as long as they are relativistic, and the free-streaming scale is nearly equal to the Hubble horizon. Therefore, the free-streaming effect suppresses perturbations below such scales.

Neutrino masses from cosmology have been studied by combining observations of CMB anisotropies with Galaxy Clustering \cite{saito2008,saito2009,saito2011}, Weak Lensing \cite{ichiki2009}, and the Lyman-$\alpha$ Forest \cite{viel2006,seljak2006}, and so on. The Planck CMB temperature power spectrum with WMAP polarization constrains the sum of the neutrino masses to be $\sum m_{\nu}<$ 0.933 eV (95\% C.L.) \cite{planck2013_16}. By combining the Planck temperature data with WMAP polarization and the high-resolution CMB data and the distance measurements from the baryon acoustic oscillations (BAO), an robust upper bound of $\sum m_{\nu} <$ 0.230 eV has been reported \cite{planck2013_16}.
In Ref. \cite{oyama2013}, by focusing on ongoing and future observations of both the 21-cm line and the CMB B-mode polarization, sensitivities to the effective number of neutrino species, total neutrino mass, and neutrino mass hierarchy are studied.

In this {\it letter}, as the most effective means of using the effect of massive neutrinos in cosmology, Planck data \cite{planck2013_1,planck2013_15} from ongoing CMB observations is used, including the B-mode polarization from CMB lensing, and the BigBOSS experiment \cite{schlegel2011,schlegel2009} is adopted for future observations of BAO and the galaxy power spectrum with an all-sky galaxy redshift survey. By comparing the observational data with models, the errors in the bounds on the total neutrino mass $\Delta\sum m_{\nu}$ are accurately estimated.


Here a flat $\Lambda$CDM model with two additional parameters of the total neutrino mass $\sum m_{\nu}$ and the effective number of neutrino species $N_\nu$ is assumed. The fiducial values of the parameters are listed in Table \ref{parameters}. For the fiducial value, we adopt the value of the maximum likelihood parameters obtained from the Planck temperature data with WMAP polarization at low multipoles \cite{planck2013_16} for the six parameters other than $\sum m_{\nu}$ and $N_\nu$. For the latter two, standard values are adopted. The neutrino mass is related to the neutrino density parameter by $\Omega_{\nu}h^2 = \Sigma m_\nu/(93.04~{\rm eV})$. The CMB temperature is taken to be $T_{\rm CMB} = 2.7255~{\rm K}$ \cite{fixsen2009}. The primordial helium fraction $Y_{\rm P}$ is a function of $\Omega_{\rm b}h^2$ and $N_{\nu}$ using the Big Bang nucleosynthesis consistency condition \cite{ichikawa2006,hamann2008}.

\begin{table}[h!]
\begin{tabular}{l c}
\hline
\hline
Parameter~ & Fiducial value \\
\hline
$\Omega_{\rm b}h^2$ & 0.02205 \\
$\Omega_{\rm c}h^2$ & 0.1199 \\
$100\theta_{\rm MC}$ & 1.04131 \\
$\tau$ & 0.089 \\
$n_{\rm s}$ & 0.9603 \\
$\ln{(10^{10}A_{\rm s})}$ & 3.089 \\
$\Sigma m_\nu$ & 0.06 \\
$N_{\nu}$ & 3.046 \\
\hline
\hline
\end{tabular}
\caption{Fiducial values of the parameters. Here $\Omega_{\rm b}h^2$ is the baryon density today. $\Omega_{\rm c}h^2$ is the cold dark matter density today, $100\theta_{\rm MC}$ is 100 $\times$ approximation to $r_*/D_{\rm A}$ (CosmoMC), $\tau$ is the Thomson scattering optical depth due to reionization, $n_{\rm s}$ is the scalar spectrum power-law index, $\ln{(10^{10}A_{\rm s})}$ is the log power of the primordial curvature perturbations, $\Sigma m_\nu$ is the sum of the neutrino masses in eV, and $N_{\nu}$ is the effective number of neutrino-like relativistic degrees of freedom. \label{parameters}}
\end{table}


Planck is the third CMB observation satellite, following COBE and WMAP. It is possible to take all of the information in the CMB temperature anisotropies, to measure the polarization of the CMB anisotropies to high accuracy. Planck also provides the thermal history of the universe during the formation of the first stars and galaxies. Polarization measurements may detect the signature of gravitational waves generated during inflation \cite{planck2006}.

This letter uses data from Planck for CMB observations including B-mode polarization due to CMB lensing. The satellite was launched in May 2009. In March 2013, initial cosmology results based on the first 15.5 months of operations were released \cite{planck2013_1,planck2013_2,planck2013_3,planck2013_4,planck2013_5,planck2013_6,planck2013_7,planck2013_8,planck2013_9,planck2013_10,planck2013_12,planck2013_13,planck2013_14,planck2013_15,planck2013_16,planck2013_17,planck2013_18,planck2013_19,planck2013_20,planck2013_21,planck2013_22,planck2013_23,planck2013_24,planck2013_25,planck2013_26,planck2013_27,planck2013_28,planck2013_29}, with analysis of temperature data but not of detailed polarization results. Detailed polarization data are scheduled to be released in 2014. Here we use mock data of the polarization of the CMB anisotropies generated by the code FuturCMB \cite{perotto2006}. Experimental specifications assumed in the computation are summarized in Table \ref{planck}.

\begin{table}[h!]
\begin{ruledtabular}
\begin{tabular}{c c c c c c}
Experiment & $f_{\rm sky}$ & $\nu$ & $\theta_{\rm FWHM}$ & $\Delta_{\rm T}$ & $\Delta_{\rm P}$ \\
 & & $[{\rm GHz}]$ & $[']$ & $[\mu{\rm K}]$ & $[\mu{\rm K}]$ \\
\hline
Planck \cite{planck2013_1} & 0.73 & 28.4 & 33.16 &  9.2 & 6.2  \\
       &      & 44.1 & 28.09 & 12.5 & 9.2  \\
       &      & 70.4 & 13.08 & 23.2 & 13.9 \\
       &      & 100  & 9.59  & 11   & 10.9 \\ 
       &      & 143  & 7.18  &  6   & 11.4 \\
       &      & 217  & 4.87  & 12   & 26.7 \\
       &      & 353  & 4.7   & 43   & 81.2 \\
\end{tabular}
\end{ruledtabular}
\caption{Experimental specifications of the CMB projects. Here $f_{\rm sky}$ is the observed fraction of the sky, $\nu$ is the observation frequency, $\theta_{\rm FWHM}$ is the angular resolution defined as the Full-Width at Half-Maximum, $\Delta_{\rm T}$ is the temperature sensitivity per pixel, and $\Delta_{\rm P}$ is the polarization sensitivity per pixel. \label{planck}}
\end{table}

BigBOSS \cite{schlegel2011,schlegel2009} is a Stage IV ground-based experiment to probe BAO and the growth of large-scale structures with a wide-area galaxy and quasar redshift survey. The experiment is designed to map the large scale structure of the universe. The resulting 3D sky atlas will contain signatures from primordial BAO that set ``standard ruler'' distance scales. Using the BAO signature, BigBOSS will measure the cosmological distance scale to better than 1\% accuracy, revealing the expansion history and growth of structure in the universe at the time when dark energy began to dominate. Co-moving volume and number of galaxies for BigBOSS is an order of magnitude larger than those for BOSS, and will measure the Hubble parameter $H(z)$ and angular diameter distance $d_A$ to $<$ 1\% accuracy. The BigBOSS BAO experiment will provide a distinct improvement in the Dark Energy Task Force figure of merit over all previous Stage III BAO experiments combined. As a precision cosmological probe, Dark Energy Science will be enhanced, and the neutrino mass and inflation will be studied.

In this letter, for observations of BAO distance scales and the galaxy power spectrum, we adopt BigBOSS parameters. We make mock data of the BAO distance scale and the galaxy power spectrum according to the projected error in the BAO distance scale and the galaxy power spectrum from BigBOSS shown in Fig. 1 of Ref. \cite{mostek2012}.


Using the above mock data, the Markov-Chain Monte-Carlo (MCMC) method \cite{lewis2002,cosmomc} is used to search cosmological parameter estimations in the multidimensional parameter space of cosmological observables. The error bounds on the cosmological parameters are estimated. Recently, the Fisher matrix has become standard for estimations of errors in cosmological parameters for future observations. However, when the phenomena are not Gaussian distributed (such as in the case of strong parameter degeneracies), the Fisher matrix formalism loses validity, as described in Ref. \cite{perotto2006}. Because all parameter likelihoods not always possible to approximate a Gaussian distribution, we use Monte Carlo simulations based on the publicly available CosmoMC code \cite{lewis2002,cosmomc} with mock observational data.

The cosmological parameter ranges to be explored with MCMC are listed in Table \ref{space}.

\begin{table}[h!]
\begin{tabular}{l c}
\hline
\hline
Parameter~ & Prior Range \\
\hline
$\Omega_{\rm b}h^2$ & 0.005 $\rightarrow$ 0.1 \\
$\Omega_{\rm c}h^2$ & 0.01 $\rightarrow$ 0.99 \\
$100\theta_{\rm MC}$ & 0.3 $\rightarrow$ 10 \\
$\tau$ & 0.01 $\rightarrow$ 0.8 \\
$n_{\rm s}$ & 0.5 $\rightarrow$ 1.5 \\
$\ln{(10^{10}A_{\rm s})}$ & 0.5 $\rightarrow$ 6.0 \\
$f_\nu$ & 0 $\rightarrow$ 1.0 \\
$N_{\nu}$ & 0.1 $\rightarrow$ 8.0 \\
\hline
\hline
\end{tabular}
\caption{The prior ranges explored in this letter. Here $f_\nu$ is the fraction of the dark matter that is in the form of massive neutrinos. Other symbols are the same as in Table \ref{parameters}. \label{space}}
\end{table}

``High accuracy default'' and ``accuracy level'' 5 are implemented in CAMB \cite{lewis2000,camb}. We use HALOFIT \cite{smith2003,takahashi2012} to include nonlinear effects in the evolution of the matter power spectrum. The chains have 1,000,000 points in CosmoMC.

\begin{figure}[h!]
\includegraphics[width=90mm]{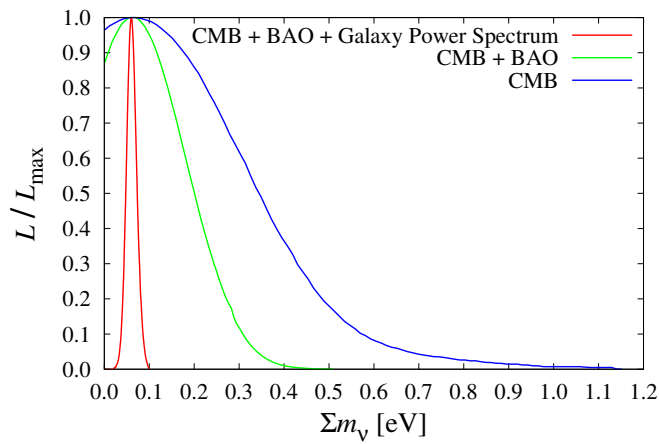}
\caption{One-dimensional probability distribution of the total neutrino mass $\sum m_{\nu}$ from observational data of CMB including B-mode (blue line), CMB including B-mode plus BAO (green line), and CMB including B-mode plus both BAO and the galaxy power spectrum (red line). 
\label{fig1}}
\end{figure}
\begin{figure}[h!]
\includegraphics[width=90mm]{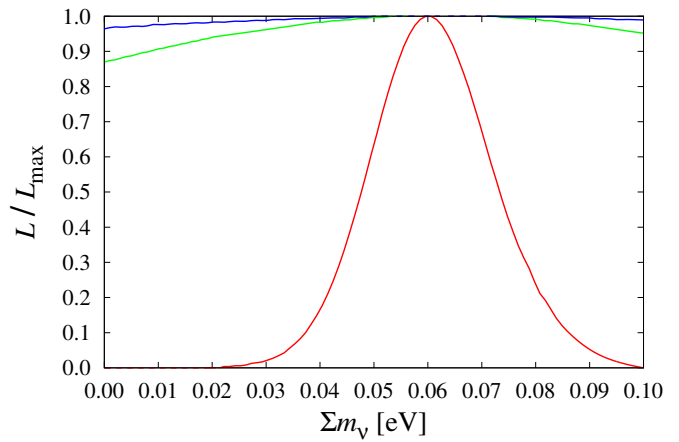}
\caption{Close-up of a portion of the CMB + BAO + Galaxy Power Spectrum from Fig \ref{fig1}. \label{fig2}}
\end{figure}

In Fig. \ref{fig1}, the probability distribution of the total neutrino mass $\sum m_{\nu}$ in eV is plotted from observational data of CMB including B-mode (Planck), with the addition of BAO (BigBOSS), and finally with the addition of both BAO and the galaxy power spectrum (BigBOSS). The fiducial value of the total neutrino mass is $\sum m_{\nu}= 0.06$ eV, whereas other parameters are marginalized. Figure \ref{fig2} enlarges the red curve from Fig. \ref{fig1}.

Using the CMB including B-mode (Planck) data, the following errors on the bounds of the total neutrino mass are obtained,
\begin{eqnarray}
\Delta\sum m_{\nu} & = & 0.24~{\rm eV~~(68\% C.L.)~~and} \nonumber \\
\Delta\sum m_{\nu} & = & 0.49~{\rm eV~~(95\% C.L.).~~~(CMB)}
\end{eqnarray}
Using the combination of the CMB including B-mode (Planck) and BAO (BigBOSS) data, the following more stringent constraint is obtained,
\begin{eqnarray}
\Delta\sum m_{\nu} & = & 0.12~{\rm eV~~(68\% C.L.)~~and} \nonumber \\
\Delta\sum m_{\nu} & = & 0.23~{\rm eV~~(95\% C.L.).~~~(CMB + BAO)}
\end{eqnarray}
Finally, using the combination of the CMB including B-mode (Planck), BAO (BigBOSS), and galaxy power spectrum (BigBOSS) data, we obtained the very stringent constraints
\begin{eqnarray}
\Delta\sum m_{\nu} & = & 0.012~{\rm eV~~(68\% C.L.) ~~and} \nonumber \\
\Delta\sum m_{\nu} & = & 0.024~{\rm eV~~(95\% C.L.).} \nonumber \\
{\rm (CMB + BAO} & {\rm +} & {\rm Galaxy~Power~Spectrum)}
\end{eqnarray}
The parameter degeneracies are efficiently broken by adding the CMB including B-mode (Planck) data to the BAO and the galaxy power spectrum (BigBOSS) data.

In summary, from the combination of the ongoing CMB observations, including the B-mode polarization due to CMB lensing (Planck), and the future observations of the BAO distance scale and galaxy power spectrum (BigBOSS), the error in the bound of the total neutrino mass is estimated to be $\Delta\sum m_{\nu} = 0.012$ eV with a 68\% confidence level. This prediction is the most stringent bound ever, portending accurate determination of neutrino masses.

Our results of the errors of the bounds are slightly tighter than the estimations using Fisher matrix in Refs. \cite{schlegel2011} and \cite{slosar2012}. The Monte Carlo simulations tend to predict slightly tighter bound on the parameters than the case of the Fisher matrix formalism, as seen in Table 1. of Ref. \cite{perotto2006}.

It is known that the total neutrino mass is $\sum m_{\nu} \hspace{0.3em}\raisebox{0.4ex}{$>$}\hspace{-0.75em}\raisebox{-.7ex}{$\sim$}\hspace{0.3em} 0.1$ eV in case of an inverted hierarchy. Hence, if the fiducial value of the total neutrino mass is $\sum m_{\nu}= 0.06$ eV, as seen in Fig. \ref{fig2}, our result implies that the neutrino mass hierarchy must be normal.

\begin{acknowledgments}
This work is supported by a Grant-in-Aid for Scientific Research from the Japan Society for the Promotion of Science (Grant Number 25400264 (KH)).
\end{acknowledgments}

\bibliography{bibliography}

\end{document}